\documentclass[aps,groupedaddress,showpacs,prl,letterpaper,twocolumn]{revtex4}
\usepackage{graphicx} % standard LaTeX graphics tool
\usepackage{amsmath}

\begin{document}

\title{Collapse and stable self-trapping for  Bose-Einstein condensates with   $1/r^b$ type attractive interatomic interaction potential}
\author{ Pavel M. Lushnikov}
\affiliation{
Department of Mathematics and Statistics,
    University of New Mexico,
    Albuquerque, New Mexico 87131
}
 \date{\today}

\begin{abstract}
We consider dynamics of Bose-Einstein condensates with long-range attractive interaction proportional to $1/r^b$ and arbitrary angular dependence.
It is shown exactly that collapse of Bose-Einstein condensate without contact interactions is  possible only for $b\ge 2$.  Case $b=2$ is critical and requires
number of particles to exceed critical value to allow collapse. Critical collapse in that case is strong one trapping into collapsing region a finite number of particles.
 Case $b>2$ is supercritical with expected weak collapse which traps rapidly decreasing number of particles during approach to collapse.  For $b<2$ singularity at $r=0$ is not strong enough to allow collapse but attractive  $1/r^b$ interaction admits stable self-trapping even in absence of external trapping potential.
 % Particular applications include dipole-dipole interaction potential, electromagnetically induced $1/r$ interaction as well as wide range of experimentally adjustable  interaction potentials between atoms in highly excited Rydberg states.
\end{abstract}

\pacs{03.75.Kk,  03.75.Lm}
\maketitle

The dynamics of   Bose-Einstein condensate (BEC) with short-range s-wave interaction have been the subject of extensive research in recent years \cite{DalfovoGiorginiPitaevskiiStringariRevModPhys1999,DonleyEtAlNature2001}. Condensates with a positive scattering length have a repulsive (defocusing)
nonlinearity  which stabilizes the condensate with the help of external trap. Condensates with a negative scattering length have an attractive (focusing) nonlinearity which formally admits solitons. However, without trap these solitons are unstable and their perturbation leads either to collapse of condensate or condensate expansion. External trap prevents expansion of condensate and makes  solitons   metastable  for a sufficiently small number of atoms. Otherwise, for larger number of atoms, the focusing nonlinearity results in  collapse of  solitons. The effect of a long-range dipolar interaction
on BEC was first studied theoretically \cite{YiYouPRA2000,Goral2000EtAlPRA2000,SantosShlyapnikovZollerLewensteinPRL2000,LushnikovPRA2002,CuevasMalomedKevrekidisFrantzeskakisPRA2009} and more recently observed experimentally \cite{GriesmaierWernerHenslerStuhlerPfauPRL2005,LahayeKochFroehlichFattoriMetzGriesmaieGiovanazziPfauNature2007,KochLahayeMetzFroehlichGriesmaierPfauNaturePhys2008} (see also \cite{BaranovPhysRep2008,LahayeMenottiSantosLewensteinPfauRepProgrPhys2009}  for review). In particular, collapse of BEC with dominant dipole-dipole forces  predicted based on approximate variational estimate  \cite{SantosShlyapnikovZollerLewensteinPRL2000}  and  obtained based on exact analysis \cite{LushnikovPRA2002} was recently observed in experiment \cite{LahayeMetzPfauEtAlPRL2008}.

Here we look for possibility of collapse of BEC due to long-range attraction vs. formation of stable self-trapped condensate for a general type of long-range interaction
\begin{equation} \label{Vdef}
V({\bf r})=\frac{f({\bf n})}{r^b}, \quad b>0, \quad {\bf n}\equiv \frac{{\bf r}}{r}, \quad r\equiv |{\bf r}|,
\end{equation}
where $f({\bf n})$ is an arbitrary bounded function  $|f({\bf n})|<\infty$ and ${\bf r}=(x_1,x_2,x_3)$. We do not require $f({\bf n})$ to be sign-definite. By attractive interaction we mean that $f({\bf n})$ is negative at least for some nonzero range of angles so that one can choose a wave function to provide negative contribution to energy functional.

Possible experimental realization of (\ref{Vdef}) are numerous. E.g., recent experimental advances allow to study interaction of ultracold Rydberg atoms with principle quantum number about 100 (see e.g.
       %\cite{AndersonVealeGallagherPhysRevLett1998,MourachkoComparatTomasiFiorettiNosbaumAkulinPilletPhysRevLett1998,HeidemannRaitzschBendkowskyButscherLowPfauPhysRevLett2008,UrbanJohnsonHenageIsenhowerYavuzWalkerSaffmanNaturePhys2009,SaffmaRevModPhys2009}).
\cite{AndersonVealeGallagherPhysRevLett1998,SaffmaRevModPhys2009}).  These interactions between atoms in highly excited Rydberg levels are long-range and dominated by dipole-dipole-type forces. Strength of interaction between Rb atoms is about $10^{12}$ times stronger (at typical distance $\sim 10 \mu m$) than interaction between Rb atoms in ground state (see e.g. \cite{SaffmaRevModPhys2009} for review). Strength and angular dependence of interaction between Rydberg atoms can be tuned in a wide range \cite{BuchlerMicheliZollerNaturePhys2007,SaffmaRevModPhys2009}. E.g., spatial dependence for Rb with principle quantum number about 100 can be $\propto 1/r^3$ for $r\lesssim 9.5 \mu m$ and $\propto 1/r^6$ (van der Waals character)  for $r\gtrsim 9.5 \mu m$ \cite{SaffmaRevModPhys2009}. Short-range s-wave scattering interaction is limited to much smaller distance $\sim$ few nm so that the range of dominance of long-range interaction potential is quite high. Another possible form of long-range attractive interaction is gravity-like $1/r$ potential which is proposed to be realized in a system of atoms with laser induced dipoles such that an
arrangement of several laser fields causes cancelation of anisotropic terms \cite{ODellGiovanazziKurizkiAkulinPRL2000}.  Terms $\propto 1/r^2$ are also possible \cite{ODellGiovanazziKurizkiAkulinPRL2000}.

The mean field BEC dynamics %with the potential  (\ref{Vdef}) 
is governed by a nonlocal Gross-Pitaevskii equation (NGPE)
\begin{eqnarray}
i\hbar\frac{\partial \Psi({\bf r})}{\partial t}&=&\left[-\frac{\hbar^2}{2m}\nabla^2+\frac{1}{2}m\omega_0^2(x_1^2+x_2^2+\gamma^2 x_3^2)
\right.\nonumber\\
&&\left.\hspace{-1.cm}
+g|\Psi({\bf r})|^2 +\int d^3{\bf r}'\, V({\bf r}-{\bf r}')|\Psi({\bf r}')|^2\right] \Psi({\bf r}),
\label{GP1}
\end{eqnarray}
where $\Psi$ is the condensate wave function,
the contact interaction is $\propto g=4\pi \hbar^2 a/m$, $a$ is the $s-$wave scattering length, $m$ is the atomic mass,
$\omega_0$ is the external  trap frequency in the $x_1-x_2$ plane,
 $\gamma$ is the anisotropy factor of the trap, and the wavefunction is normalized to the number of atoms, $\int d^3{\bf r}\, |\Psi|^2=N$.
 Contact interaction term can be also included into potential $V({\bf r})$ as  $\frac{g}{2}\delta({\bf r})$ but we have not done that because we focus here on effect of long-rage potential (\ref{Vdef}).
 If $V({\bf r})\equiv 0$ then a standard Gross-Pitaevski equation (GPE) \cite{DalfovoGiorginiPitaevskiiStringariRevModPhys1999} is recovered.

NGPE $(\ref{GP1})$ can be  written through variation $  i\hbar\frac{\partial\Psi}{\partial t}=
\frac{\delta E}{\delta \Psi^*}$ of the energy
functional %$E:$ , where the  condensate energy,
\begin{eqnarray}\label{Edef}
E=E_{K}+E_P+E_{NL}+E_{R},
\end{eqnarray}
which is an integral of motion: $\frac{ d \,E}{d \,t}=0$, and
\begin{equation}\label{Etotdef}
\begin{split}
 E_K&= \int\frac{\hbar^2}{2m} |\nabla \Psi|^2 d^3{\bf r}, \quad E_{NL}=\frac{g}{2}\int |\Psi|^4 d^3{\bf r},
\\
E_P&=\int \frac{1}{2} m\omega_0^2(x_1^2+x_2^2+
\gamma^2x_3^2)|\Psi|^2d^3{\bf r},
\\
E_{R}&=\frac{1}{2}\int |\Psi({\bf r})|^2 V({\bf r}-{\bf
r}')|\Psi({\bf r}')|^2 d^3{\bf r}d^3{\bf r}'.
\end{split}
\end{equation}

Consider time evolution of the mean square radius of the wave
function, $\langle r^2 \rangle \equiv \int r^2 |\Psi|^2 d^3{\bf
r}/N.$ Using $(\ref{GP1}),$ integrating by parts, and taking into
account vanishing boundary conditions at infinity one obtains
\begin{eqnarray}\label{At}
\partial_t\langle r^2 \rangle=\frac{\hbar}{2mN} \int 2 i x_j  (\Psi\partial_{x_j}\Psi^\ast-\Psi^\ast\partial_{x_j}\Psi)d^3{\bf r},
\end{eqnarray}
where $\partial_{t}\equiv \frac{\partial}{\partial t}$,
$\partial_{x_j}\equiv \frac{\partial}{\partial x_j}$ and repeated
index $j$ means summation over all space coordinates, $j=1,
\ldots, 3$. After a second differentiation over $t$, one
gets \cite{LushnikovPRA2002}
\begin{eqnarray}\label{Att}
\partial^2_t\langle r^2 \rangle=\frac{1}{2mN}\Big [
8E_K-8E_P+12E_{NL} \quad \qquad \qquad\quad \nonumber\\
-2\int|\Psi({\bf r}|^2 |\Psi({\bf r'}|^2(x_j
\partial_{x_j}+x'_j\partial_{x'_j})V({\bf r}-{\bf r}')d^3{\bf r}\Big ],
\end{eqnarray}
which is called by a virial theorem \cite{LushnikovPRA2002} similar to
GPE
\cite{VlasovPetrishchevTalanovRdiofiz1971,ZakharovJETP1972,LushnikovJETPLett1995,PitaevskiiPhysLettA1996,BergePhysRep1998,LushnikovSaffmanPRE2000}.

It follows from $(\ref{Vdef})$ that $(x_j
\partial_{x_j}+x'_j\partial_{x'_j})V({\bf r}-{\bf r}')=-bV({\bf r}-{\bf
r}')$
%we use that (x_j \partial_{x_j}+x'_j\partial_{x'_j})=_j \partial_{p_j}+q_j\partial_{q_j}, \quad p_j\equiv(x_j+x_j'),  \quad q_j\equiv(x_j-x_j')
 and using $(\ref{Edef})$ we rewrite $(\ref{Att})$ as follows 
\begin{eqnarray}\label{Att2}
\partial^2_t\langle r^2 \rangle=\frac{1}{2mN}\Big
[4bE+(8-4b)E_K-(4+2b)m\omega_0^2N\langle r^2\rangle
\nonumber\\
-(4+2b)m\omega_0^2N(\gamma^2-1)\langle x_3^2\rangle+(12-4b)E_{NL}\Big ].
\end{eqnarray}
Here  both the local nonlinear term $E_{NL}$ and the nonlocal nonlinear term $E_R$ are included into the energy
$E$.
 %which is a conserved quantity. 
Catastrophic collapse of BEC in terms of NGPE means a singularity formation, $\max|\Psi|\to \infty$, in a finite time. Because of conservation of $N$, the typical size of atomic cloud near singularity must vanish.
The virial theorem
$(\ref{Att2})$ describes collapse when  the positive-definite quantity $\langle r^2\rangle$ becomes negative in finite
time implying $\max|\Psi|\to \infty$   before $\langle
r^2\rangle$ turns negative. The kinetic energy $E_K$ diverges at collapse time which follows from divergence of the potential energy for $\max|\Psi|\to \infty$ together with
conservation of the energy functional $E$.
Another way to see divergence of $E_K$ is from uncertainty relation $E_K\ge \frac{\hbar^2}{2m}   (9/4)N/\langle r^2\rangle$ (see \cite{LushnikovJETPLett1995,LushnikovPRA2002})
for $\langle r^2\rangle\to 0$.
 Generally $\langle r^2\rangle$ may not vanish at collapse (e.g. if there are nonzero values of $|\Psi|$ away from collapse center) but
$E_K$  diverges at collapse time for sure because of $\max|\Psi|\to \infty$. We  use below divergence of $E_K$ as  necessary and sufficient condition of collapse formation while vanishing of
$\langle r^2\rangle$ is only sufficient condition for collapse.

NGPE is not applicable  near singularity  and
another physical mechanisms are important such as inelastic two- and three-body collisions which
can cause a loss of atoms from the condensate \cite{DalfovoGiorginiPitaevskiiStringariRevModPhys1999}. In addition, multipole expansion used for derivation
of the dipole-dipole-type potential is not applicable on a very short distances (few  a few Bohr radii). However, as explained above, NGPE with potential (\ref{Vdef}) is a good approximation  for a wide range of typical interatomic distances.

Consider case $2\le b \le 3$. Then one immediately obtains from
equation $(\ref{Att2})$ that $\partial^2_t\langle r^2
\rangle\le\frac{6E}{mN}$. Integrating that differential inequality over time we get that $\langle r^2
\rangle\le\frac{3E}{mN}t^2+\partial_t\langle r^2
\rangle|_{t=0} t+\langle r^2 \rangle|_{t=0}$. If $E<0$ we  conclude that
 $\langle r^2\rangle \to 0$ for large enough $t$ which provides a sufficient
criterion of collapse of BEC.
Condition $E<0$ is sufficient but not necessary for collapse. Using generalized uncertainty relations between $E_K, \
N, \langle r^2\rangle, \ \partial _t\langle r^2\rangle$ \cite{LushnikovJETPLett1995,LushnikovPRA2002} one can obtain much stricter condition of collapse which is  outside the scope of this Letter.

Below we assume  $g=0$. Choosing e.g. initial condition as $\psi|_{t=0}=\frac{N^{1/2}}{\pi^{3/4}r_0^{3/2}}e^{-r^2/(2r_0^2)}$  gives $E=-\frac{3\hbar^2}{4m}\frac{N}{r_0^2}+\pi^{-1/2}2^{4-b}f N^2r_0^{-b}\Gamma(3/2-b/2)$ for $\omega_0=0$ and  $f({\bf n})=Const\equiv f.$ If the constant  $r_0$ is small and $b>2$ or if $N>3\hbar^2/(16m f)$ and $b=2$ then $E<0$.
It means that for $2\le b \le 3$ long-distance potential alone is enough to achieve catastrophic collapse of BEC. Another particular example was considered in Ref. \cite{LushnikovPRA2002}
for the case of dipole-dipole interaction potential with all dipoles oriented in one fixed direction. Note also that $b=3$ is a border between short-range potentials (for $b>3$) and long-range potentials (for $b\le 3$) in 3D
(one needs convergence of  $\int_{r<r_c} V({\bf r}) d^3{\bf r}$  to have short-range potential, where $r_c$ is a cutoff at small distances). Case $b=3$   also requires that integral of  $V({\bf r})$ over angles to be zero to ensure convergence of (\ref{Vdef}) at small $r$ (as is the case \cite{LushnikovPRA2002}  for dipole-dipole interactions with fixed direction) otherwise we would have to introduce cutoff at small distances and potential would loose general form  (\ref{Vdef}).

Case $b>3$ generally requires introduction of cutoff at small scales. Because $\int_{r<r_c} V({\bf r}) d^3{\bf r}$ is finite in that case we generally have situation very similar to standard $\delta$-correlated potential \cite{DalfovoGiorginiPitaevskiiStringariRevModPhys1999}.

\begin{figure}%[htbp]
\begin{center}
\includegraphics[width = 3.0 in]{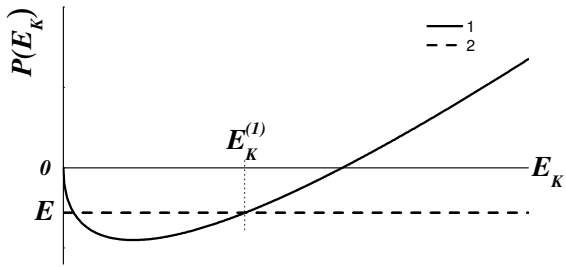}
%\vspace{0.5cm}
\caption{Schematic of the function $P(E_K)$ defined in (\ref{Etotinequal})  (solid curve). Equation $P(E_K)=E$ ($E$ corresponds to dashed line) has either one of two roots for $E_K>0$
depending on sign of $E$.  $E_K^{(1)}$ is the largest root.} \label{fig:fig1}
\end{center}
\end{figure}

Now we prove that for $b<2$ collapse is impossible for $g=0$ because singularity of (\ref{Vdef}) is not strong enough. We use the inequality 
%
%\begin{equation*}\label{ladyzhinequal}
$\int\frac{|\Psi({\bf r})|^2}{|{\bf r}-{\bf r}'|^2} d^3{\bf r}\le 4 \int{|\nabla \Psi({\bf r})|^2} d^3{\bf r}$ %\end{equation*}
\cite{Ladyzhenskaya1969}, which holds for any ${\bf r}'$. Using now H\"older's inequality
we generalize that inequality (assuming $b<2$) as follows
%
%\begin{equation*}%\label{ladyzhinequalgeneralization}
%\begin{split}
%&\int\frac{|\Psi({\bf r})|^2}{|{\bf r}-{\bf r}'|^b} d^3{\bf r}=\int|\Psi({\bf r})|^{2-b}\frac{|\Psi({\bf r})|^b}{|{\bf r}-{\bf r}'|^b} d^3{\bf r}
%\\
%&\le \left [\int\left (|\Psi({\bf r})|^{2-b}\right )^\frac{1}{1-b/2} d^3{\bf r}\right ]^{1-\frac{b}{2}}   
%\end{split}
%\end{equation*}
%
%
\begin{equation}\label{ladyzhinequalgeneralization}
\begin{split}
&\int\frac{|\Psi({\bf r})|^2}{|{\bf r}-{\bf r}'|^b} d^3{\bf r}=\int|\Psi({\bf r})|^{2-b}\frac{|\Psi({\bf r})|^b}{|{\bf r}-{\bf r}'|^b} d^3{\bf r}
\\
&\le \left [\int\left (|\Psi({\bf r})|^{2-b}\right )^\frac{1}{1-b/2} d^3{\bf r}\right ]^{1-\frac{b}{2}}
\\
   &\times  \left [\int\left (\frac{|\Psi({\bf r})|^b}{|{\bf r}-{\bf r}'|^b}\right )^{\frac{2}{b}}\right ]^{\frac{b}{2}} d^3{\bf r}         \le 2^b N^{1-\frac{b}{2}} \left (\frac{2m}{\hbar^2}E_K\right )^{\frac{b}{2}}.
\end{split}
\end{equation}
Using now boundness of $f:$ $f({\bf n})\le f_m\equiv \max\limits_{\bf n} |f({\bf n})|$ in (\ref{Vdef})  and inequality (\ref{ladyzhinequalgeneralization}) we obtain a bound for $E_R$ in (\ref{Etotdef})
\begin{equation}\label{ERinequal}
\begin{split}
 E_R& \ge -f_m\, 2^{b-1} N^{2-\frac{b}{2}} \left (\frac{2m}{\hbar^2}E_K\right )^{\frac{b}{2}},
\end{split}
\end{equation}
which gives a respective bound of $E$ in  (\ref{Etotdef}) (recall that we assume $g=0$):
\begin{equation}\label{Etotinequal}
\begin{split}
 E& \ge E_K-f_m\, 2^{b-1} N^{2-\frac{b}{2}} \left (\frac{2m}{\hbar^2}E_K\right )^{\frac{b}{2}}\equiv P(E_K).
\end{split}
\end{equation}
A function $P(E_K)$ in   (\ref{Etotinequal}) has a minimum for $E_K\equiv E_K^{(0)}=2^{-2}[f_m\,  b]^{2/(2-b)}  N^{\frac{4-b}{2-b}}(2m/{\hbar^2})^{b/(2-b)}$ resulting in a lower bound
\begin{equation}\label{Etotinequallower}
\begin{split}
 E& \ge -\frac{2-b}{b}2^{-2}[f_m\,  b]^{\frac{2}{2-b}}  N^{\frac{4-b}{2-b}}\left (\frac{2m}{\hbar^2}\right )^{\frac{b}{2-b}}.
\end{split}
\end{equation}
Boundness of the energy functional $E$ from below ensures that collapse is impossible for $b<2$. To prove  that  we show  that  $E_K$ is bounded while collapse requires $E_K\to \infty$.
 We choose any value of $E$ which satisfy (\ref{Etotinequallower}). Fig. \ref{fig:fig1} shows schematically  the function $P(E_K)$ from (\ref{Etotinequal}).
 %It has a minimum at $E_K= E_K^{(0)}$ and
 Inequality (\ref{Etotinequal}) requires that $E_K\le E_K^{(1)}(E)$, where $E_K^{(1)}(E)$ is the largest root or equation $P(E_K)=E$.  It proves that $E_K$ is bounded for fixed $N$ which completes the proof of absence of collapse for $b<2$. Particular version of that result for $b=1$ and $f({\bf n})=const$ was first obtained in Ref. \cite{TuritsynTheorMathPhys1985}. Nonexistence of collapse  for nonsingular potential  $V({\bf r})$ was shown previously  based on approximate analysis in Ref. \cite{PerezKonotopPRE2000}. Proof of nonexistence of collapse for particular example  of nonsingular potentials with positive-definite bounded Fourier transform was given in Ref. \cite{BangKrolikowskiWyllerRasmussenPRE2002}. These results can be easily generalized for any bounded potential similar to above analysis. Thus collapse can occur for singular potential only and singularity should be strong enough, i.e. $b\ge 2.$

%%%%%%%%%%%%%%%%%%%%%%

We now look for soliton solution of NGPE (\ref{GP1}) as $\Psi({\bf r},t)=A({\bf r})e^{-i\mu t/\hbar }$, where $\mu$ is the chemical potential.
In that case NGPE  (\ref{GP1}) reduces to time-independent equation
\begin{eqnarray}
&&\left[-\mu-\frac{\hbar^2}{2m}\nabla^2+\frac{1}{2}m\omega_0^2(x_1^2+x_2^2+\gamma^2 x_3^2)
\right.\nonumber\\
&&\left.\hspace{-1.cm}
\qquad \qquad+\int d^3{\bf r}'\, V({\bf r}-{\bf r}')A({\bf r}')^2\right] A({\bf r})=0,
\label{GP1Asoliton}
\end{eqnarray}
where we again assume $g=0$ although generalization to $g\ne 0$ case is straightforward. Equation (\ref{GP1Asoliton}) is the stationary point of the energy functional $E$ for a fixed number of particles: $\delta (E-\mu N)=0.$ Multiplying equation (\ref{GP1Asoliton}) by $A$ and $x_j\partial_{x_j}A$ and integrating by parts one
obtains using (\ref{Vdef}) and (\ref{Edef}) that
\begin{equation}\label{XYHs}
\begin{split}
 E_{K,s}&=-\mu N_s\frac{b}{4-b}+E_{P,s}, \quad  E_{R,s}=\mu N_s\frac{2}{4-b},
\\
E_{s}&=-\mu N_s\frac{b-2}{4-b}+2E_{P,s},
\end{split}
\end{equation}
 where subscript "s" means values of all integrals on soliton solution. Especially simple and interesting is the case of self-trapping ($\omega_0=0$) when condensate is in steady state without any external trap.
 All integrals in that case depend on number of particles $ N_s$ only.

 Assume radial symmetry $f({\bf n})=Const<0$ in (\ref{Vdef}). Ground state soliton is determined from condition that $A({\bf r})$ never crosses zero \cite{ZakharovKuznetsovJETP1974rus,KuznetsovRubenchikZakharovPhysRep1986}. To prove ground soliton stability we show that
 it realizes a minimum of the Hamiltonian for a fixed $N_s.$  One can make inequality (\ref{ladyzhinequalgeneralization}) sharper by minimizing a functional ${\cal F}(\Psi)\equiv N^{1-\frac{b}{2}} \left (\frac{2m}{\hbar^2}E_K\right )^{\frac{b}{2}}/\int\frac{|\Psi({\bf r})|^2}{|{\bf r}-{\bf r}'|^b} d^3{\bf r}$. That minimum is achieved at one of stationary points $\frac{\delta{\cal F}}{\delta \Psi^*}=0$ and after simple rescaling one can see that these points corresponds to  soliton solutions of the time-independent NGPE (\ref{GP1Asoliton}). Among these stationary points the minimum is achieved at ground state soliton $\Psi_{s,ground}$. It gives a bound ${\cal F}(\Psi)\ge\min{\cal F}(\Psi)={\cal F}(\Psi_{s,ground})$ which is sharper than the inequality (\ref{ladyzhinequalgeneralization}). Most important is that following analysis similar to equations  (\ref{ERinequal})-(\ref{Etotinequallower}) we obtain that for any $\Psi$
\begin{equation}\label{EtotinequallowerESground}
\begin{split}
 E& \ge \min E=E_{s,ground},
\end{split}
\end{equation}
i.e. the ground state soliton solution attains the minimum of $E$ for fixed $N$. % which can be written as a stationary point of variational problem ...
It proves exactly the stability of soliton for $f({\bf n})=Const$. Similar ideas were  used in a  nonlinear Schr\"odinger equation                
% (GPE  with $\omega_0=0$). 
\cite{ZakharovKuznetsovJETP1974rus,KuznetsovRubenchikZakharovPhysRep1986}. Ground state soliton was also found numerically for $b=1$ \cite{CartariusFabcicMainWunnerPRA2008}.

For more general  $f({\bf n})\ne Const$ minimum of $E$ is still negative if $f({\bf n})$ is negative for a nonzero range of values of $\bf n$. So in that case we expect that ground state soliton solution attains that minimum and, respectively is stable. If $f({\bf n})>0$ for any $\bf n$ then $\min E=0$. It corresponds to unbounded spatial spreading of NGPE solution for any initial conditions. Self-trapping is impossible in that case and  solitons are possible for  $\omega_0\ne 0$ only.

Case $b=2$ is on the boundary between  bounded and unbounded energy functional as can be seen from inequalities (\ref{Etotinequal}) and (\ref{EtotinequallowerESground}).
If $N>N_{s,ground}$ then $E$ is unbounded. If  $N<N_{s,ground}$ then $E\ge E_{s,ground}=0$ as follows from  (\ref{XYHs}) for $b=2.$ Thus $N_{s,ground}$ is the critical number of particles for collapse. That critical number is similar to collapse of  GPE  in dimension 2 (2D) (as well as similar to critical power in nonlinear optics) \cite{VlasovPetrishchevTalanovRdiofiz1971}. It is important to distinguish that critical number from the critical number of particles of  3D GPE with $\omega_0\ne 0$  \cite{DalfovoGiorginiPitaevskiiStringariRevModPhys1999,DonleyEtAlNature2001}. % which is basically a condition for stability of weakly nonlinear solution of standard GPE in that trap while w

To qualitatively distinguish different regimes of collapse and solitons one can consider, in addition to the exact analysis above, a scaling transformations \cite{ZakharovKuznetsovJETP1986rus} which conserves the number of particles $\Psi({\bf r})\to a^{-3/2}\Psi({\bf r}/a)$. Under this transformation the energy functional $E$ (for $\omega_0=0$) depends on the parameter $a$ as follows
\begin{equation}\label{Escaling}
E(a)=a^{-2}E_K+a^{-b}E_R.
\end{equation}

\begin{figure}%[htbp]
\begin{center}
\includegraphics[width = 3.0 in]{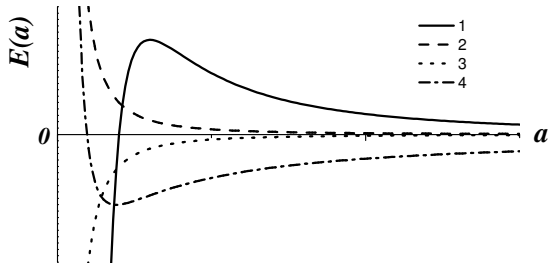}
%\vspace{0.5cm}
\caption{ Schematic of $E(a)$ from  (\ref{Escaling}) for   $b>2$, $b=2$ and $b<2$ (see text for details).  } \label{fig:fig2}
\end{center}
\end{figure}

The virial theorem (\ref{Att2}) and the relations (\ref{XYHs}), (\ref{Escaling}) have striking similarities with  GPE if we replace $b$ by the spatial dimension $D$ in  GPE. That analogy suggests to refer the case $b=2$  as the critical NGPE and $b>2$ as the supercritical NGPE. Fig. \ref{fig:fig2} shows typical dependence of (\ref{Escaling}) on $a$ for $b>2$, $b=2$ and $b<2$ assuming $E_R<0$.
For $b>2$ there is a maximum of $E$ (curve 1 in Fig. \ref{fig:fig2}) corresponding to unstable soliton. Solution of NGPE either collapse or expand.   For  $b=2$  there is no extremum and
collapse is impossible for $N<N_{s,ground}$  (curve 2 in Fig. \ref{fig:fig2})   while condensate can collapse for  $N>N_{s,ground}$   (curve 3 in Fig. \ref{fig:fig2}). Ground state soliton corresponds to $N=N_{s,ground}$ and $E=0$ locating exactly at the boundary between collapsing and noncollapsing regimes.
For $b<2$ there is a   minimum which corresponds to stable ground state soliton  (curve 4 in Fig. \ref{fig:fig2}).

Solutions of both GPE and NGPE  with $\omega_0=0$  near collapse typically consist of background of nearly linear waves and a central collapsing self-similar nonlinear core.  The scaling (\ref{Escaling}) describes the dynamics of the core with time-dependent $a$ such that $a\to 0$ near collapse.   Waves have negligible potential energy but carry a positive kinetic energy $E_{waves}\simeq E_{K,waves}$. The total energy  $E=E_{collapse}+E_{waves}$ is constant, where $E_{collapse}$ is the  core energy.
It follows from  (\ref{Escaling}) that for  $b=2$  one can simultaneously allow conservation of $N$ and $E_{collapse}$ so that  negligible number of  waves are emitted from the core. This  scenario is called a strong collapse.  If $b>2$  then the term $\propto a^{-b}$ in  (\ref{Escaling}) dominates with $E_{collapse}\to -\infty$ as $a\to 0$. Then the only way to ensure conservation of $E$ is to assume strong emission of linear waves.  Near collapse time only vanishing number of particles remains in a core (of course all that is true until NGPE losses its applicability) which is called weak collapse \cite{ZakharovKuznetsovJETP1986rus}. Strong collapse was shown to be unstable for supercritical GPE \cite{ZakharovKuznetsovJETP1986rus}.
It suggests that collapse for $b>2$ is of weak type.

Support was provided by NSF grant DMS 
 0807131.

\end{document}